# Femtosecond Transfer and Manipulation of Persistent Hot-Trion Coherence in a Single CdSe/ZnSe Quantum Dot

P. Henzler[1], C. Traum[1], M. Holtkemper[2], D. Nabben[1], M. Erbe[1], D. E. Reiter[2], T. Kuhn[2], S. Mahapatra[3,†], K. Brunner[3], D. V. Seletskiy[1,4], and A. Leitenstorfer[1,*]

[1] *Department of Physics and Center for Applied Photonics, University of Konstanz, D-78457 Konstanz, Germany*
[2] *Institute of Solid State Theory, University of Münster, D-48149 Münster, Germany*
[3] *Institute of Physics, EP3, University of Würzburg, D-97074 Würzburg, Germany*
[4] *Department of Engineering Physics, Polytechnique Montréal, Montréal, Québec H3T 1J4, Canada*
† *Current affiliation: Department of Physics, Indian Institute of Technology Bombay, Mumbai, India*
\* *Author e-mail address: Alfred.Leitenstorfer@uni-konstanz.de*

**Abstract:**

Ultrafast transmission changes around the fundamental trion resonance are studied after exciting a $p$-shell exciton in a negatively charged II-VI quantum dot. The biexcitonic induced absorption reveals quantum beats between hot-trion states at 133 GHz. While interband dephasing is dominated by relaxation of the $P$-shell hole within 390 fs, trionic coherence remains stored in the spin system for 85 ps due to Pauli blocking of the triplet electron. The complex spectrotemporal evolution of transmission is explained analytically by solving the Maxwell-Liouville equations. Pump and probe polarizations provide full control over amplitude and phase of the quantum beats.

Technological advances based on genuine quantum phenomena combine multiple opportunities and challenges. In general, the coherence time is a crucial parameter [1,2]. Therefore, understanding intrinsic relaxation and dephasing mechanisms in elementary quantum systems is key to further progress [3–5]. Long-lived coherences are typically assigned to electronic states close to equilibrium, where protection from pure dephasing is well known [6–10]. Despite the importance of highly excited states for quantum technology [11,12], their relaxation and dephasing dynamics remains poorly understood. In this context, the restricted phase space and large transition dipoles in semiconductor quantum dots (QDs) [13] offer interesting perspectives for spin-to-photon interfaces [2,14–17]. Specimens based on II-VI compounds may be especially advantageous since strong electronic confinement and Coulomb interactions enhance energy separations [18], enabling coherent manipulation even with femtosecond light pulses [14]. In principle, driving such quantum systems far from equilibrium allows to study both lifetime and potential transfer of quantum coherence between excited states as well as full relaxation pathways of individual charge carriers.

In this Letter, we report ultrafast generation and manipulation of a persistent coherence between excited trion states of a single negatively-charged CdSe/ZnSe QD. Spectral changes of induced absorption into biexcitonic states directly reveal quantum beats between trion triplet states. Very surprisingly, the coherence between hot-trion states is transferred upon scattering of the photoexcited hole within 390 fs. Subsequently, it remains protected by the Pauli blocking of hot-electron relaxation and persists for 85 ps, i.e., almost three orders of magnitude longer than the timescale required for coherently initializing and manipulating the quantum system.

Our experimental setup consists of a three-color femtosecond fiber source coupled to a polarization-sensitive transmission microscope operating at 1.6 K [19]. Individual CdSe/ZnSe QDs [20,21] are embedded into subwavelength Al apertures to increase light-matter coupling [14,19]. Interband excitation generates a trion comprising two electrons and one hole. FIG. 1(a) shows a micro-photoluminescence spectrum. At an energy of 2.1482 eV, the radiative recombination $X^-$ of the trion ground state ($|TGS\rangle$) into the global ground state ($|QDGS\rangle$) is observed. Two emission lines $XX_X^-$ and $XX_Y^-$ appear redshifted to $X^-$ at energies of 2.1429 eV and 2.1434 eV, respectively. A quadratic increase of intensity with excitation power [21] assigns them to recombination of the charged biexciton ground state ($|CBGS\rangle$) into trion triplet states $|X\rangle$ and $|Y\rangle$ [22] which are spectrally split by (550±5) µeV. Note that we work with an excitation intensity weak enough to ensure a low probability for generation of a biexciton by the pump. A level scheme together with relevant electronic configurations is depicted in FIG. 1(b). $|X\rangle$ and $|Y\rangle$ are both composed of one hole in the $S$ shell of the valence band and two electrons, one each in the conduction-band $s$ and $p$ shells. $|X^*\rangle$ and $|Y^*\rangle$ have identical electron configurations but the hole occupies the $P$ shell. Various spin configurations split in energy by exchange interactions emerge [14,22–25]. Specifically, the electron-electron exchange lifts the degeneracy between singlet and triplet configurations. Electron-hole exchange then separates the triplet states into two bright levels $|X\rangle$ and $|Y\rangle$, depending on the in-plane asymmetry of the confinement potential. Close to cylindrical symmetry, $XX_X^-$ and $XX_Y^-$ are circularly polarized. $|X\rangle$ and $|Y\rangle$ then decay into the $|TGS\rangle$ with distinctly different relaxation times [14]. A significant deviation from cylindrical symmetry results in a linear polarization of $XX_X^-$ and $XX_Y^-$ [22] along the principal axes $\vec{e}_X$ and $\vec{e}_Y$ of the confinement potential and similar relaxation rates of $|X\rangle$ and $|Y\rangle$, respectively. Polarization-sensitive photoluminescence (PL) spectroscopy assigns our QD to the latter type [21].

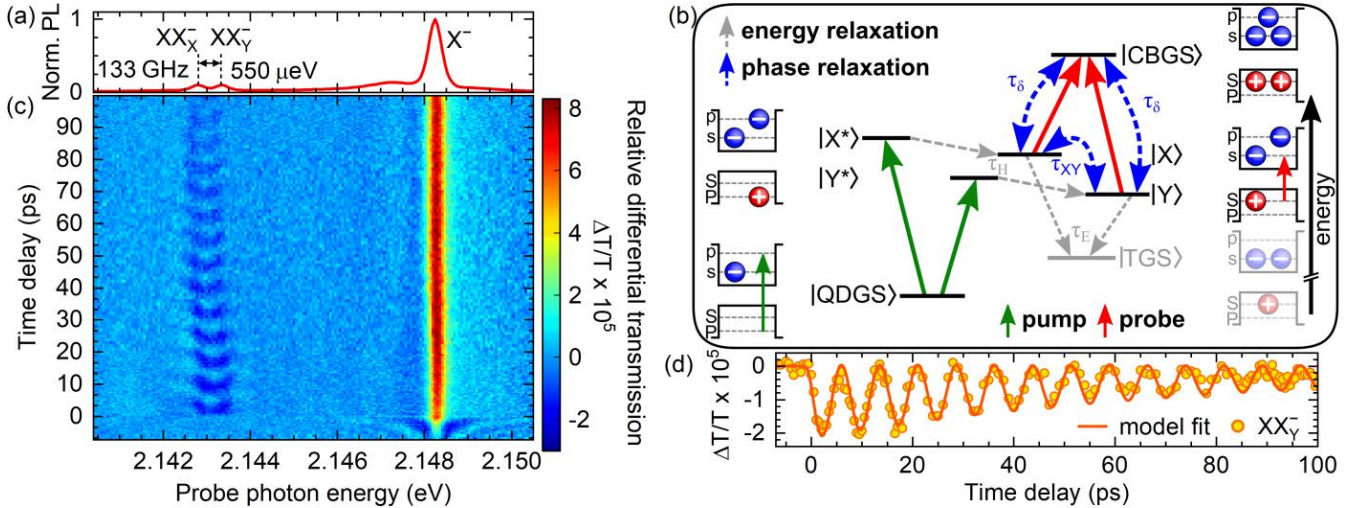

FIG. 1. Relative differential transmission of a single CdSe/ZnSe QD. Pulses polarized linearly along $\vec{e}_X - \vec{e}_Y$ initialize a superposition of excited states $|X^*\rangle$ and $|Y^*\rangle$, probed collinearly. (a) Micro-photoluminescence. $X^-$ denotes the fundamental trion resonance. $XX_X^-$ and $XX_Y^-$ indicate recombination of the biexciton ground state $|CBGS\rangle$ into triplet states $|X\rangle$ or $|Y\rangle$. (b) Few-level system. Radiative and non-radiative transitions are marked with solid (green and red for pumping and probing) and dashed gray arrows, respectively. Dashed blue arrows indicate phase relationships between eigenstates. $\tau_H$ marks the hole relaxation time, $\tau_E$ the $p$-shell electron relaxation time, $\tau_\delta$ the interband dephasing time and $\tau_{XY}$ the dephasing time of the quantum beats. (c) Relative differential transmission $\Delta T/T$ color-coded versus photon energy and pump-probe delay time. (d) Dynamics of $\Delta T/T$ around $XX_Y^-$, visualized as yellow-orange circles with an error margin of $\pm 2 \times 10^{-6}$. The orange line represents a least-square fit to our theory.

We now excite $p$-shell transitions of the QD initializing a coherent superposition of $|X^*\rangle$ and $|Y^*\rangle$ with 520-fs pulses of a central photon energy of 2.228 eV and spectral width of 5 meV, which are linearly polarized along $\vec{e}_X - \vec{e}_Y$ (green arrows in FIG. 1(b)). 100-fs probe pulses are collinearly polarized and centered at 2.145 eV. Their bandwidth of 25 meV covers the entire range of fundamental trion and biexciton emission. Typical average powers for incident excitation and readout pulse trains are 10 µW and 1 µW, corresponding to pulse areas of 0.72 $\pi$ and 0.22 $\pi$, respectively [21]. FIG. 1(c) shows the color-coded relative differential transmission $\Delta T/T$ as a function of photon energy and time delay $t_D$ between pumping and probing. Where probe pulses precede excitation at negative $t_D$, delay-dependent modulations at X$^-$ result from a perturbed free induction decay [15,26]. For positive $t_D$, two processes contribute equally to the signal at X$^-$. First, ultrafast bleaching due to Coulomb renormalization [14,15] results in a steep increase to half of the maximum $\Delta T/T$ on a timescale below 1 ps [21]. Subsequently, single-photon gain emerges on a 100-ps timescale when population inversion between |TGS⟩ and |QDGS⟩ is established, directly revealing the intraband scattering times from |X*⟩ and |Y*⟩ into |TGS⟩ [14]. As we will show below, the timescale for establishing the |TGS⟩ is completely dominated by electron relaxation because the scattering of the hole, i.e. the step from |X*⟩ or |Y*⟩ to |X⟩ or |Y⟩, respectively, proceeds very rapidly. A |TGS⟩ recombination time $\tau_{|TGS⟩}$ of (366±33) ps is deduced from $\Delta T/T$ at even longer delays. At the energy of XX$_X^-$ and XX$_Y^-$, negative signatures appear for positive $t_D$. They originate from activating optical transitions from |X⟩ and |Y⟩ into |CBGS⟩, as indicated by red arrows in FIG. 1(b). The most striking feature in this region is a long-lived periodic modulation of the line shape of biexcitonic induced absorption. A fast Fourier transform of $\Delta T/T$ at XX$_Y^-$ reveals an oscillation frequency of (133±2) GHz, coinciding exactly with the energy difference between the biexcitonic emission lines of (550±5) μeV. This finding suggests that the signal emerges from quantum beats between |X⟩ and |Y⟩, as indicated by a dashed blue arrow and dephasing time $\tau_{XY}$ in FIG. 1(b). The modulation is analyzed in more detail in FIG. 1(d). The yellow-orange circles result from spectrally integrating $\Delta T/T$ around the position of XX$_Y^-$ within an interval of 0.4 meV. The decay of the amplitude is caused by the relaxation of the $p$-shell electron to the $s$ shell, corresponding to the transitions from |X⟩ and |Y⟩ into |TGS⟩. From a model fit to the data in FIG. 1(d), we extract a time constant of $\tau_E = (85±10)$ ps for this process (see gray dashed arrows in FIG 1(b)). The consistency of our picture is underlined by the rise time of single-photon gain at X$^-$ (FIG. 1(c)) of (83±12) ps [21]: scattering of electrons from the $p$ shell directly populates the $s$ shell, thus establishing the |TGS⟩ with a time constant identical to $\tau_E$. Compared to other excited trion states [14], the lifetime of |X⟩ and |Y⟩ is one to two orders of magnitude longer and merely a factor of five shorter than the interband recombination time $\tau_{|TGS⟩}$. These unusual conditions are due to Pauli blocking by the resident electron: relaxation of a triplet electron requires a combined electron-hole spin flip [14,24], rendering these states metastable. While energy relaxation is encoded in the signal envelope in FIG. 1(d), the trion coherence manifests itself in the contrast of the underlying oscillations. Interestingly, the quantum beats are clearly present and even persist over the entire temporal range of finite amplitude of biexcitonic absorption. This finding indicates that the coherence between |X*⟩ and |Y*⟩ is conserved during the relaxation into |X⟩ and |Y⟩ and even remains protected from pure dephasing: the decay time $\tau_{XY}$ of (85±10) ps we extract from the oscillation contrast again coincides with $\tau_E$. Obviously, the coherence is limited exclusively by the population relaxation of |X⟩ and |Y⟩, requiring a combined electron-hole spin-flip. Note that $\tau_{XY}$ is much larger than the interband dephasing time $\tau_\delta$ of (3.7±0.5) ps between |X⟩ ↔ |CBGS⟩ and |Y⟩ ↔ |CBGS⟩ derived from the PL linewidth of XX$_X^-$ and XX$_Y^-$ of (360±30) μeV which is not limited by our spectral resolution of 100 μeV [19].

Considering the carrier-phonon interaction within a Lindblad model provides a microscopic understanding of both the conservation of coherence during relaxation of the hole and the absence of pure dephasing thereafter [21,27,28]. The essential point is that the electron-phonon coupling acts solely on the orbital part of an electronic wave function. Both |X⟩ and |Y⟩ as well as |X*⟩ and |Y*⟩ share the same orbital state and only differ in their spin configuration. Thus, on the one hand, a pure relaxation of the orbital part of the hole from $P$ to $S$ shell does not affect the spin coherence between |X⟩ and |Y⟩ or |X*⟩ and |Y*⟩). On the other hand, all phonon scattering processes between coherent superpositions of |X⟩ and |Y⟩ or |X*⟩ and |Y*⟩ are strongly correlated, thus preventing pure dephasing.

To analyze the lineshape modulation of biexcitonic signatures in FIG. 1(c), we calculate $\Delta T/T$ using the Maxwell-Liouville equations. The polarization of the QD acts as a source for a re-emitted field which is superimposed with the much stronger probe field, forming the total transmitted electric field [29]. For an analytical solution, we restrict ourselves to linearly polarized transitions and assume $\delta(t)$-shaped pulses [21]. The result coincides with a numerical solution including realistic light pulses and states based on a configuration interaction approach [23]. Adopting small probe intensities and identical transition dipoles from $|QDGS\rangle$ into $|X^*\rangle$ and $|Y^*\rangle$ as well as from $|X\rangle$ and $|Y\rangle$ into $|CBGS\rangle$ [30,31], we find

$$(\Delta T/T)_{X/Y} \sim \frac{-1/\tau_\delta}{1/\tau_\delta^2 + (\omega_{X/Y} - \omega)^2} \cdot \left( \underbrace{e^{-t_D/\tau_E}}_{(A)} + \underbrace{e^{-t_D/\tau_{XY}} \cos(\omega_{XY} t_D + \vartheta)}_{(B)} \right) \mp \frac{\omega_{X/Y} - \omega}{1/\tau_\delta^2 + (\omega_{X/Y} - \omega)^2} \cdot \underbrace{e^{-t_D/\tau_{XY}} \sin(\omega_{XY} t_D + \vartheta)}_{(C)} \quad (1)$$

where $\omega_{X/Y}$ correspond to the photon frequencies of transitions $XX^-_{X/Y}$ and $\omega_{XY} = \omega_Y - \omega_X$. As discussed below, the relative polarization of excitation and readout defines the phase $\vartheta$ of the oscillations. The total differential transmission is $\Delta T/T = (\Delta T/T)_X + (\Delta T/T)_Y$. Three major contributions evident from the right-hand side of Eq. (1) are visualized in FIG. 2.

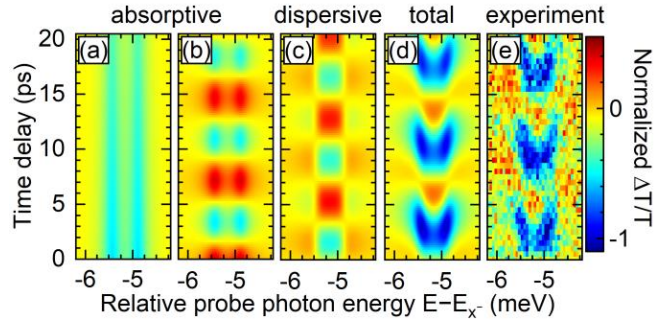

FIG. 2. Temporal evolution of analytical and experimental $\Delta T/T$ around the biexcitonic absorption color-coded versus photon energy relative to the $X^-$ transition at 2.1482 eV. Calculated static-absorptive (a), modulated-absorptive (b) and dispersive contributions (c) add up to the total signal (d). The corresponding subset of experimental data from FIG. 1(c) is depicted in (e).

The first part of the solution referring to term (A) in Eq. (1) is depicted in FIG. 2(a). It even occurs for an incoherent occupation of $|X\rangle$ and $|Y\rangle$ and is described by a Lorentzian-shaped absorption for each transition without any temporal modulation. The second (B) and third (C) parts of Eq. (1) are related to the excitonic coherence. Consequently, they oscillate with $\omega_{XY}$ and decay with $e^{-t_D/\tau_{XY}}$. Part (B) is visualized in FIG. 2(b). It corresponds to a periodic modulation of the statically induced absorption in FIG. 2(a) with the beating frequency $\omega_{XY}$. Interestingly, a third component (C) arises (see FIG. 2(c)), which is phase shifted by $\pi/2$ with respect to the direct modulation of absorption (FIG. 2(b)): the non-stationary evolution of the electronic states causes a phase modulation of the re-emitted field [21], creating new frequency components in regions of maximum temporal change. The full evolution of transient transmission is obtained by summing all three contributions (A) to (C), as shown in FIG. 2(d). The V-shaped forms (blue) and positive regions (dark orange) represent an excellent match to the experimental results in FIG. 2(e).

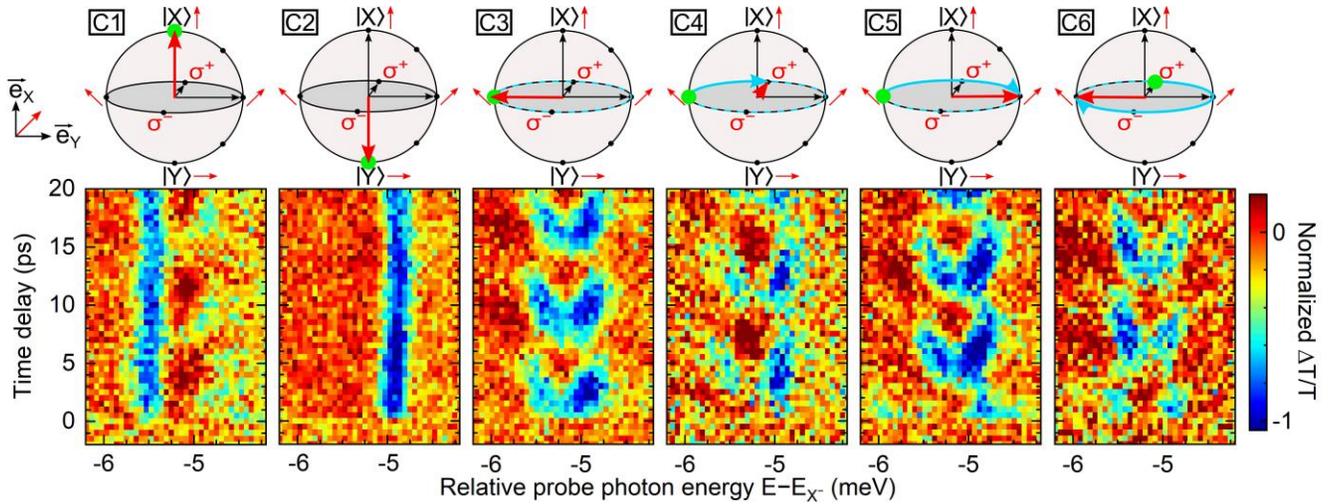

FIG. 3. Polarization control of quantum beats. Top: polarization configurations C1 to C6 visualized on Bloch spheres, representing superpositions of $|X\rangle$ and $|Y\rangle$. Points on the sphere are associated with probe polarizations (small red arrows and $\sigma^{+,-}$) relative to the axes $\vec{e}_X$ and $\vec{e}_Y$ (coordinate system at left). Green dots mark the excited superposition, thick red arrows the read-out one. If a modulation exists (C3-C6), it is visualized in light blue (dashed line and arrow), rotating clockwise around the equator. Polarization configurations: C1 pump-probe linear along $\vec{e}_X$; C2 pump-probe linear $\vec{e}_Y$; C3 pump-probe linear $\vec{e}_X - \vec{e}_Y$; C4 pump linear $\vec{e}_X - \vec{e}_Y$, probe right-circular; C5 pump linear $\vec{e}_X - \vec{e}_Y$, probe linear $\vec{e}_X + \vec{e}_Y$; C6 pump left-circular, probe linear $\vec{e}_X - \vec{e}_Y$. Bottom: transient transmission for C1 to C6 color-coded versus time delay and probe photon energy relative to $X^-$.

We now control the quantum beats by varying pump-probe polarizations. Six different configurations C1 to C6 are visualized in Bloch spheres at the top of FIG. 3, representing coherent superpositions of states $|X\rangle$ and $|Y\rangle$. Linear probe polarizations are visualized by thin red arrows. Their direction refers to the principal axes $\vec{e}_X$ and $\vec{e}_Y$ of the confinement potential, as exemplified by the coordinate system at left. Circular probe polarizations are marked by $\sigma^{+,-}$. Coherent superpositions initialized by the pump are indicated by green dots. Note that here, $\sigma^+$ and $\sigma^-$ must be interchanged for pump and probe due to the selection rules for two-step resonant biexciton excitations. Each specific probe polarization is indicated by a thick red arrow pointing towards the readout state.

For C1 and C2, excitation and readout are set collinearly along $\vec{e}_X$ or $\vec{e}_Y$ to exclusively excite and probe $|X\rangle$ or $|Y\rangle$. The differential transmissions (FIG. 3) indeed show biexcitonic absorption solely at $XX_X^-$ (C1) and at $XX_Y^-$ (C2), respectively. No modulation occurs due to excitation of an eigenstate. In C3 to C6, coherent superpositions with identical contributions from $|X\rangle$ and $|Y\rangle$ are initialized. A collinear polarization along $\vec{e}_X - \vec{e}_Y$ is excited in C3, corresponding to the configuration discussed in FIG. 1 and 2. We now control the phase of the coherent beats by changing the probe polarization. C4 and C5 exhibit phase shifts of $\pi/2$ and $\pi$, respectively. Their origin is visualized by light blue arrows in the Bloch spheres: after initializing a specific superposition, the maximum amplitude is reached when it rotates in the equatorial plane until it phases up with the probe. Note that solely the relative polarization angle between pump and probe determines the temporal phase. This fact is demonstrated by C6 with a circular polarization used for pumping: a probe polarization identical to C3 shifts the phase by $3\pi/2$.

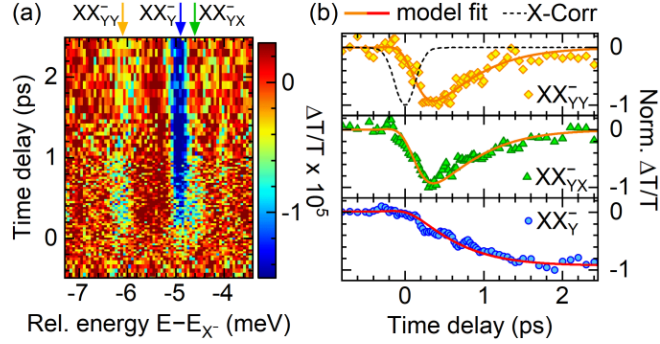

FIG. 4. Femtosecond relaxation of *P*-shell hole. (a) Color-coded differential probe transmission versus time delay and photon energy relative to X⁻, as measured in configuration C2. (b) Spectral slices of normalized $\Delta T/T$ integrated over an energy interval of 0.4 meV centered around -6.1 meV ($XX^-_{YY}$), -4.9 meV ($XX^-_{Y}$) and -4.6 meV ($XX^-_{YX}$) versus time delay. Solid orange and red lines: least-square fits to theoretical model. Dashed black line: intensity cross-correlation between pump and probe.

We can now understand why these signatures emerge so rapidly after excitation, despite the required relaxation from $|X^*\rangle$ and $|Y^*\rangle$ to $|X\rangle$ and $|Y\rangle$. Interestingly, only a broad resonance with a width of (3.4±0.4) meV is discernible in a photoluminescence excitation (PLE) spectrum of the QD [21] where transitions from $|QDGS\rangle$ into $|X^*\rangle$ and $|Y^*\rangle$ are located. Naively, one would expect two sharp resonances enabling the formation of a long-lived superposition state. These facts indicate an ultrafast relaxation of the *P*-shell hole into the *S* shell, providing both a rapid onset of biexcitonic absorption and a fast interband dephasing. For direct experimental access to the hole scattering, we measure with configuration C2 to isolate the dynamics leading from $|Y^*\rangle$ to $|Y\rangle$ (FIG. 4(a)). Here, three transient absorptions emerge early after excitation with only one of them located at $XX^-_Y$ (indicated by the blue arrow). Centered 6.1 meV below X⁻ (energy position marked with orange arrow), a second signature $XX^-_{YY}$ appears redshifted to both biexcitonic resonances. A third feature $XX^-_{YX}$ is slightly blue-shifted to and partially overlaps with $XX^-_Y$. Its energetic position is centered 4.6 meV below X⁻ and marked with a green arrow. Note that as $XX^-_{YY}$, also feature $XX^-_{YX}$ is confined to short times $t_D$ between 0 ps and 1 ps. We assign both lines to an induced absorption, establishing a hot biexciton including one *P*-shell hole: while residing in the photoexcited state $|Y^*\rangle$ (FIG. 1(b)), absorbing a photon with an energy close to $XX^-_Y$ forms a hot biexciton. The minute shifts of $XX^-_{YY}$ and $XX^-_{YX}$ with respect to $XX^-_Y$ result from slightly different Coulomb energies of the other involved carriers with respect to the *P*-shell hole. Spectral slices integrated over a 0.4-meV interval around $XX^-_{YY}$ and $XX^-_{YX}$ are depicted as yellow diamonds and green triangles in FIG. 4(b), respectively. The orange lines result from modelling the increase of induced absorption with the convolution of a non-resonant cross-correlation between pump and probe (dashed line at top of FIG. 4(b)) and an exponential decay related to a hole relaxation time $\tau_H$ of (390±80) fs. As expected, the dynamics match for $XX^-_{YY}$ and $XX^-_{YX}$. If the increase of absorption at $XX^-_Y$ originates from relaxation of the *P*-shell hole, it should occur with the same time constant. Indeed, the time integral over the fitting functions at $XX^-_{YY}$ and $XX^-_{YX}$ (red graph at the bottom of FIG. 4(b)) agrees well with the blue circles extracted by spectrally slicing data from FIG. 4(a) around $XX^-_Y$. By PLE measurements, we determine the energy gap between valence-band *S* and *D* shells to (60±10) meV [21]. Assuming parabolic confinement, a gap of (30±5) meV between *P* and *S* shell results which is close to the longitudinal-optical phonon energy of CdSe of 25 meV [32]. This fact explains the femtosecond hole relaxation with a quasiresonant quantum kinetic coupling via the Fröhlich interaction [14,33].

In summary, we find a subnanosecond coherence time between hot-trion states in a single QD. Control over amplitude and phase of the resulting quantum beats in biexcitonic absorption is provided by the pump-probe polarizations. This option allows us to directly investigate the femtosecond relaxation of the *P*-shell hole. In contrast, trion spin coherence is protected from dephasing by identical orbital shell configurations of the states and limited solely by the energy relaxation of the *p*-shell electron requiring an electron-hole spin flip.

This combination results in a difference between interband dephasing and trion coherence times by almost three orders of magnitude. The significant amount of fine-structure splitting between hot-trion states provides sub-THz frequencies for the evolution of coherent superpositions. We expect analogous phenomena to occur also in other species of zero-dimensional quantum systems whenever electron relaxation is slowed, e.g., by Pauli blocking and exchange splitting is large enough to provide adequate beat frequencies. These facts are rendering our observations relevant for the search of promising platforms for ultrafast quantum logic operations with extremely large processing bandwidth.

Funding by the DFG via collaborative research center SFB767 is gratefully acknowledged.

## References:

[1]    D. P. DiVincenzo and D. Loss, J. Magn. Magn. Mater. **200**, 202 (1999).

[2]    P. Michler, *Quantum Dots for Quantum Information Technologies* (Springer International Publishing, Cham, 2017).

[3]    A. B. Henriques, A. Schwan, S. Varwig, A. D. B. Maia, A. A. Quivy, D. R. Yakovlev, and M. Bayer, Phys. Rev. B **86**, 115333 (2012).

[4]    O. Gazzano, S. Michaelis de Vasconcellos, C. Arnold, A. Nowak, E. Galopin, I. Sagnes, L. Lanco, A. Lemaître, and P. Senellart, Nat. Commun. **4**, 1425 (2013).

[5]    S. Sim, D. Lee, A. V. Trifonov, T. Kim, S. Cha, J. H. Sung, S. Cho, W. Shim, M.-H. Jo, and H. Choi, Nat. Commun. **9**, 351 (2018).

[6]    J. Berezovsky, M. H. Mikkelsen, N. G. Stoltz, L. A. Coldren, and D. D. Awschalom, Science **320**, 349 (2008).

[7]    G. Moody, R. Singh, H. Li, I. A. Akimov, M. Bayer, D. Reuter, A. D. Wieck, and S. T. Cundiff, Solid State Commun. **163**, 65 (2013).

[8]    V. Giesz, N. Somaschi, G. Hornecker, T. Grange, B. Reznychenko, L. De Santis, J. Demory, C. Gomez, I. Sagnes, A. Lemaître, O. Krebs, N. D. Lanzillotti-Kimura, L. Lanco, A. Auffeves, and P. Senellart, Nat. Commun. **7**, 11986 (2016).

[9]    D. Ding, M. H. Appel, A. Javadi, X. Zhou, M. C. Löbl, I. Söllner, R. Schott, C. Papon, T. Pregnolato, L. Midolo, A. D. Wieck, A. Ludwig, R. J. Warburton, T. Schröder, and P. Lodahl, Phys. Rev. Appl. **11**, 031002 (2019).

[10]   M. Geller, Appl. Phys. Rev. **6**, 031306 (2019).

[11]   I. Schwartz, E. R. Schmidgall, L. Gantz, D. Cogan, E. Bordo, Y. Don, M. Zielinski, and D. Gershoni, Phys. Rev. X **5**, 011009 (2015).

[12]   T. Suzuki, R. Singh, G. Moody, M. Aßmann, M. Bayer, A. Ludwig, A. D. Wieck, and S. T. Cundiff, Phys. Rev. B **98**, 195304 (2018).

[13]   U. Woggon, *Optical Properties of Semiconductor Quantum Dots* (Springer, 1997).

[14]   C. Hinz, P. Gumbsheimer, C. Traum, M. Holtkemper, B. Bauer, J. Haase, S. Mahapatra, A. Frey, K. Brunner, D. E. Reiter, T. Kuhn, D. V. Seletskiy, and A. Leitenstorfer, Phys. Rev. B **97**, 045302 (2018).

[15]   F. Sotier, T. Thomay, T. Hanke, J. Korger, S. Mahapatra, A. Frey, K. Brunner, R. Bratschitsch, and A. Leitenstorfer, Nat. Physics **5**, 352 (2009).

[16] A. Greilich, S. E. Economou, S. Spatzek, D. R. Yakovlev, D. Reuter, A. D. Wieck, T. L. Reinecke, and M. Bayer, Nat. Physics **5**, 262 (2009).

[17] D. Press, T. D. Ladd, B. Zhang, and Y. Yamamoto, Nature **456**, 218 (2008).

[18] O. Zakharov, A. Rubio, X. Blase, M. L. Cohen, and S. G. Louie, Phys. Rev. B **50**, 10780 (1994).

[19] C. Traum, P. Henzler, S. Lohner, H. Becker, D. Nabben, P. Gumbsheimer, C. Hinz, J. F. Lippmann, S. Mahapatra, K. Brunner, D. V. Seletskiy, and A. Leitenstorfer, Rev. Sci. Instrum. **90**, 123003 (2019).

[20] S. Mahapatra, K. Brunner, and C. Bougerol, Appl. Phys. Lett. **91**, 153110 (2007).

[21] See Supplemental Material below for details concerning sample structures and stationary characterization, the dynamics of the fundamental trion resonance as well as derivations of spin configurations of trion triplet states and of Eq. (1).

[22] I. A. Akimov, K. V. Kavokin, A. Hundt, and F. Henneberger, Phys. Rev. B **71**, 075326 (2005).

[23] M. Holtkemper, D. E. Reiter, and T. Kuhn, Phys. Rev. B **97**, 075308 (2018).

[24] J. Huneke, I. D'Amico, P. Machnikowski, T. Thomay, R. Bratschitsch, A. Leitenstorfer, and T. Kuhn, Phys. Rev. B **84**, 115320 (2011).

[25] M. E. Ware, E. A. Stinaff, D. Gammon, M. F. Doty, A. S. Bracker, D. Gershoni, V. L. Korenev, Ş. C. Bădescu, Y. Lyanda-Geller, and T. L. Reinecke, Phys. Rev. Lett. **95**, 177403 (2005).

[26] T. Guenther, C. Lienau, T. Elsaesser, M. Glanemann, V. M. Axt, T. Kuhn, S. Eshlaghi, and A. D. Wieck, Phys. Rev. Lett. **89**, 057401 (2002).

[27] F. Rossi and T. Kuhn, Rev. Mod. Phys. **74**, 895 (2002).

[28] D. E. Reiter, T. Kuhn, and V. M. Axt, Adv. Phys. X **4**, 1655478 (2019).

[29] S. Mukamel, *Principles of Nonlinear Optical Spectroscopy* (Oxford University Press, 1995).

[30] D. E. Reiter, Phys. Rev. B **95**, 125308 (2017).

[31] D. E. Reiter, V. M. Axt, and T. Kuhn, Phys. Rev. B **87**, 115430 (2013).

[32] C. I. Contescu and K. Putyera, *Dekker Encyclopedia of Nanoscience and Nanotechnology* (CRC Press, 2009).

[33] E. A. Zibik, T. Grange, B. A. Carpenter, N. E. Porter, R. Ferreira, G. Bastard, D. Stehr, S. Winnerl, M. Helm, H. Y. Liu, M. S. Skolnick, and L. R. Wilson, Nat. Materials **8**, 803 (2009).

# Supplemental Material

# Femtosecond Transfer and Manipulation of Persistent Hot-Trion Coherence in a Single CdSe/ZnSe Quantum Dot

P. Henzler, C. Traum, M. Holtkemper, D. Nabben, M. Erbe, D. E. Reiter, T. Kuhn,
S. Mahapatra, K. Brunner, D. V. Seletskiy, and A. Leitenstorfer

In the following, we provide additional information, supporting measurements, a full derivation of Eq. (1) and an explanation of our theoretical model, as referred to in the main paper.

## Sample Structures

In the main paper, we investigate a negatively charged CdSe/n-ZnSe quantum dot (QD) grown by molecular beam epitaxy on a GaAs (001) substrate by a Te-mediated self-assembly process [1]. Formation of well-defined 3D islands in different self-assembled patterns was observed by atomic force microscopy (AFM) of uncapped CdSe/ZnSe ensembles. However, high-resolution transmission electron microscopy (HRTEM)-based composition mapping revealed that the ZnSe-capped CdSe QDs are essentially undulations in a compositionally inhomogeneous quasi-2D CdZnSe layer with Cd-rich cores. These QDs extend laterally by up to 8 nm. They are separated by approximately 100 nm, corresponding to an areal density of $10^{10}$ cm$^{-2}$. To optimize the interaction probability between optical pulses and single charge carriers in the quantum structure, we use the nanophotonic concept [2] of embedding the QDs in sub-wavelength Al apertures.

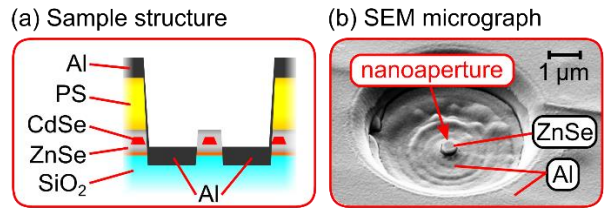

FIG.S 1: (a) Sketch of the nanophotonic sample structure. CdSe QDs (red truncated cones) in a ZnSe matrix of 100 nm thickness (light grey) are embedded in a nanoaperture formed by a 110-nm-thick layer of aluminum (Al, dark grey). A sacrificial layer of polystyrene (PS, yellow) is necessary for the preparation process. The structure is placed on a quartz substrate (SiO$_2$, light blue). (b) SEM micrograph of a sample structure with a nanoaperture featuring a diameter of approximately 250 nm.

A sketch of our sample structure is depicted in FIG.S 1(a). The CdSe QDs are shown as red truncated cones, embedded in the ZnSe matrix (light grey) of 100 nm thickness. We enclose discs of this ZnSe matrix in sub-wavelength apertures with a diameter of 250 nm by means of a focused ion beam (FIB) milling process combined with a sequence of different evaporation and lift-off steps. The nanoapertures are formed by an aluminum (Al) layer of 110 nm thickness, shown in dark grey, while the 500-nm-thick polystyrene layer underneath (PS, yellow) only serves as sacrificial layer for the preparation process. The spatial resolution of the FIB milling of less than 7 nm [2] allows for high geometrical precision in preparing the nanoapertures. Owing to their almost perfect circular shape, we can perform measurements involving a defined helicity of excitation and readout pulses. Additionally, the geometry reduces the effective beam diameter, allowing us to exploit transmission properties of sub-wavelength apertures [3]. Placing the CdSe/ZnSe layer on a quartz substrate (SiO$_2$, light blue) of 150 µm thickness before nanoaperture preparation enables transient transmission experiments. A SEM micrograph of the final sample structure is shown in FIG.S 1(b).

## Photoluminescence Spectroscopy

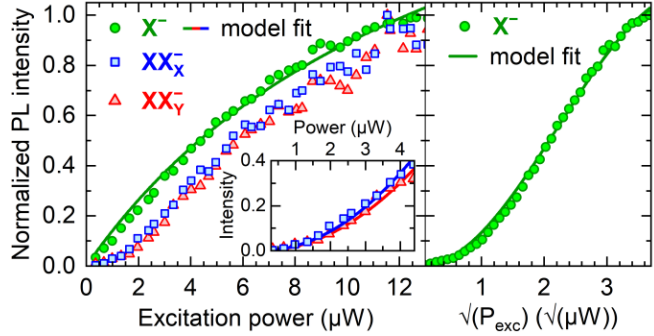

FIG.S 2: Excitation power-dependent photoluminescence emission of the QD, complementing FIG.1. Pump pulses are centered at an energy of 2.228 eV and linearly polarized along $\vec{e}_X - \vec{e}_Y$. Left panel: The normalized intensities of the X$^-$ resonance and the biexcitonic emission lines XX$_X^-$ and XX$_Y^-$ are depicted versus excitation power. The green line shows the calculated dependence for a three-level model. Inset: The normalized intensity of the biexcitonic emission increases quadratically (blue and red graphs) for low excitation powers. Right panel: The normalized intensity of the X$^-$ resonance is plotted versus pulse area, determined as the square root of the excitation power.

Excitation power-resolved PL spectroscopy reveals the origin of discernable signatures as well as the excitation pulse area. For this experiment, pump pulses are set to simultaneously excite $|X^*\rangle$ and $|Y^*\rangle$. The left panel of FIG.S 2 depicts the variation of the normalized PL intensity with the excitation power for the three relevant transitions X$^-$ (green dots), XX$_X^-$ (blue squares) and XX$_Y^-$ (red triangles). In each case the PL intensity is integrated over their corresponding spectral width of (360±30) µeV. The power dependence of the X$^-$ intensity follows the green line, calculated in accordance to a three-level-system [4] with an approximately linear dependence for low-power excitation. At the extracted saturation power of (15±1) µW, half of the maximum amplitude is reached. The inset of the left panel shows the power dependence of the biexcitonic emission lines. The quadratic increase indicates that two electron-hole pairs are excited in a resonant two-step process forming a biexciton.

The right panel of FIG.S 2 shows the normalized integrated PL intensity at the X$^-$ resonance versus the square root of the excitation power which is proportional to the electric field strength of the excitation. A least-square fit to a sine function reveals a pulse area of $\pi$ for $\sqrt{(P_{exc})} = (4.4\pm0.1) \sqrt{(\mu W)}$. Pumping with 10 µW thus corresponds to a pulse area of 0.72 $\pi$.

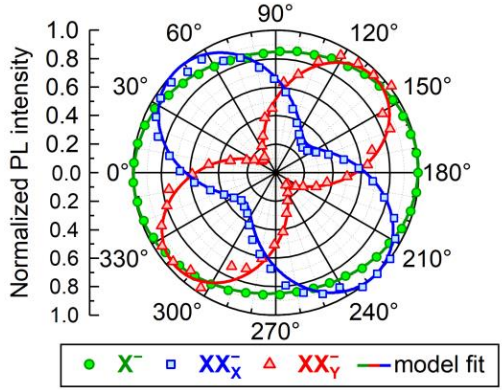

FIG.S 3: Polarization-resolved PL spectroscopy of the QD studied in FIG.1. Pump pulses simultaneously excite both states $|X^*\rangle$ and $|Y^*\rangle$. The normalized PL intensity is plotted over the doubled rotation angle of a half-wave plate, thus revealing the polarization of the PL. A least-square fit determines the degree of linear polarization and its phase.

Polarization-resolved PL spectroscopy reveals the orientation of the electronic confinement potential of the QD in the growth plane. Pump pulses are again set to excite $|X^*\rangle$ and $|Y^*\rangle$ simultaneously. FIG.S 3 shows the normalized PL intensity of $X^-$ (green dots), $XX_X^-$ (blue squares) and $XX_Y^-$ (red triangles) versus twice the rotation angle of a half-wave plate located before the detector. To quantify the degree of linear polarization $\rho$, we define $\rho = (I_{max} - I_{min})/(I_{max} + I_{min})$ with $I_{max}$ as the maximum PL intensity and $I_{min}$ as its minimum. We model the normalized intensity $I$ according to $I(\varphi) = I_{min} + (I_{max} - I_{min}) \cdot \sin^2\left[\frac{\pi}{180°}(\varphi + \varphi_0)\right]$. In this way, we obtain $\rho_{X^-} = (0.09 \pm 0.01)$, revealing the non-polarized character of the $X^-$ resonance. This behavior corresponds to the physical interpretation of a non-correlated emission of single photons and agrees with the findings in Ref. [5]. In contrast, the biexcitonic emissions are linearly polarized with $\rho_{XX_X^-} = (0.54 \pm 0.01)$ and $\rho_{XX_Y^-} = (0.82 \pm 0.01)$. The difference in offset $\varphi_0$ of $(90.0 \pm 1.1)°$ confirms their orthogonality. Again, these findings are in good agreement with Ref. [5]. From these data, we have determined the orientation of the axes $\vec{e}_X$ and $\vec{e}_Y$ of the electronic confinement potential in our experiment.

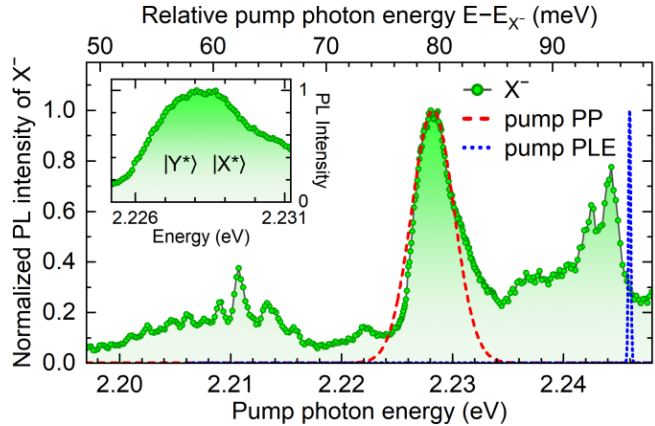

FIG.S 4: Energy-resolved PL excitation (PLE) spectroscopy of the QD studied in FIG.1. The normalized PL intensity integrated over the spectral width of the $X^-$ resonance is depicted versus pump photon energy. The intensity spectrum of the pump pulse used in the femtosecond differential transmission experiments is shown as a dashed red line. An example for the tunable excitation pulse employed for the PLE measurement is depicted as a blue dashed line, demonstrating a spectral resolution of 240 µeV.

We use energy-resolved PL excitation spectroscopy (PLE) to identify absorption resonances of the investigated QD. Theoretical calculations [6,7] assign those resonances to specific optical transitions in the QD. FIG.S 4 shows the normalized intensity of the $X^-$ transition versus the pump photon energy as green circles. Exciting the quantum system with a narrowband and tunable excitation pulse featuring a spectral width as low as 240 µeV (see blue dashed line for an example spectrum) determines the resolution of the absorption spectra [8]. The energetically lowest resonances around 2.21 eV or at (60±10) meV above $X^-$ are assigned to optical transitions into the $|Ds\rangle$ state, containing two $s$-shell electrons and one $D$-shell hole [2]. The absorption of pump photons centered around 2.228 eV drives the QD into the excited triplet states $|X^*\rangle$ and $|Y^*\rangle$. The inset shows this feature in detail. The asymmetry of the resonance is due to absorption assisted by the generation of acoustic phonons at energies above the zero-phonon line. Excluding this additional broadening, we extract a spectral width of (3.4±0.4) meV. Note that only a single broadened feature is visible in the PLE spectrum, in contrast to the two sharp lines expected from the persistent quantum beats between $|X\rangle$ and $|Y\rangle$. This finding is due to the extremely different levels of interband dephasing and coherence lifetime in our system, as discussed in the main paper. A spectrum of the 520-fs excitation pulse used in the transient transmission experiments is depicted by a dashed red line.

The absorption resonance of the singlet state, containing one $s$-shell electron, one $p$-shell electron and one $P$-shell hole appears blue-shifted to the investigated triplet absorption resonance around an energy of 2.26 eV [2,6].

# Evolution of Differential Transmission at the Fundamental Trion Resonance

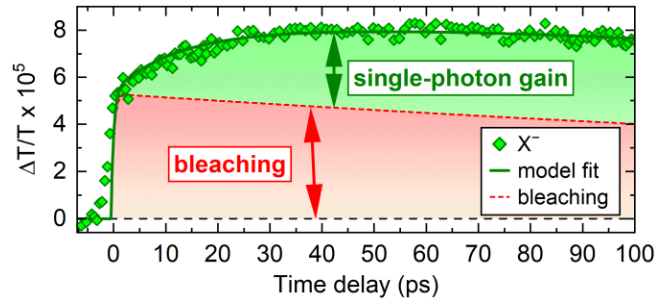

FIG.S 5 Temporal evolution of bleaching and single-photon gain at the $X^-$ resonance. Green diamonds depict a slice of the differential transmission data integrated around the energy of the $X^-$ resonance within an interval of 0.4 meV with an error margin of ±2x10$^{-6}$. This signal consists of two contributions: bleaching of the transition due to Coulomb renormalization (red shading) and single-photon gain which emerges due to inversion of the system (green shading).

For a quantitative understanding of the TGS dynamics, we analyze the differential transmission signal $\Delta T/T$ around the $X^-$ resonance, as depicted in FIG. 1(c) in the main paper. FIG.S 5 shows a slice of this data within an interval of 0.4 meV around $X^-$ as green diamonds. The signal is formed by two contributions [2]: bleaching and single-photon gain, visualized by red and green shading, respectively. The latter coincides exactly with the dynamical $|TGS\rangle$ occupation which is equivalent to optical gain. Adapting a least-square fit according to a combined model system [2] (green line) directly reveals a time constant of (83±12) ps for establishing single-photon gain in our quantum system.

# Spin Configuration of Bright Trion Triplet States

The bright $p$-shell triplet states $|X^*\rangle$ and $|Y^*\rangle$ were identified in Ref. [2] within the measured photoluminescence excitation spectra by a comparison with calculations using a configuration interaction approach based on an envelope function approximation. As described in detail in Ref. [2], the spin configurations of the two bright triplet states depend on the interplay between several interactions and thereby on details of the QD geometry. To estimate the possible spin configurations, we use an effective model including the three triplet as well as the singlet spin states (see FIG.S 6). Spin mixtures occur between singlet and the energetically lower bright triplet state ($\delta_{ee}$) as well as between the two bright triplet states ($\delta_{eh}$). The resulting realistic spin configurations of the two bright triplet states can be represented by

$$|X^*\rangle_{\chi,d} = \frac{\chi}{\sqrt{1+\chi^2}} \frac{1-d}{\sqrt{(1-d)^2+d^2}} |\downarrow\uparrow\Downarrow\rangle + \frac{1}{\sqrt{1+\chi^2}} |\downarrow\downarrow\Uparrow\rangle + \frac{\chi}{\sqrt{1+\chi^2}} \frac{d}{\sqrt{(1-d)^2+d^2}} |\uparrow\downarrow\Downarrow\rangle \quad (S\,1)$$

$$|Y^*\rangle_{\chi,d} = \frac{1}{\sqrt{1+\chi^2}} \frac{1-d}{\sqrt{(1-d)^2+d^2}} |\downarrow\uparrow\Downarrow\rangle - \frac{\chi}{\sqrt{1+\chi^2}} |\downarrow\downarrow\Uparrow\rangle + \frac{1}{\sqrt{1+\chi^2}} \frac{d}{\sqrt{(1-d)^2+d^2}} |\uparrow\downarrow\Downarrow\rangle \quad (S\,2)$$

with the two coefficients $\chi$ and $d$, associated with $\delta_{ee}$ and $\delta_{eh}$, respectively. Black (gray) ↓/↑ represent the spins of the $p$-shell ($s$-shell) electron, while ⇓/⇑ represent the heavy-hole (HH) spins. For each state $|X^*\rangle$ and $|Y^*\rangle$ exists a corresponding degenerate state with inverted spins. In $|X\rangle$ and $|Y\rangle$, similar mixtures as in $|X^*\rangle$ and $|Y^*\rangle$ are expected. For $d = 0$, values of $\chi$ define the polarization of the associated transitions between circular ($\chi = 0$) and linear ($\chi = 1$). Values of $d$ may vary between $d = 0.5$ and $d = 0$. They slightly affect the polarization and brightness of the associated transitions and are rather unknown as they are hard to access experimentally. Throughout the paper, we discuss the simplified case $\chi = 1$ and $d = 0$, where $|X^*\rangle$ and $|Y^*\rangle$ ($|X\rangle$ and $|Y\rangle$) cause purely linearly polarized transitions and the quantum beats are most pronounced and clear. Calculations treating deviations from the idealized case show effects like intensity differences between the $XX_X^-$ and $XX_Y^-$ lines (as also faintly visible in FIG. 3 C3-C6) or a non-vanishing second line in configurations like C1 and C2 (as also faintly visible in FIG. 3 C1). However, details of such deviations from the idealized case go beyond the scope of the present paper.

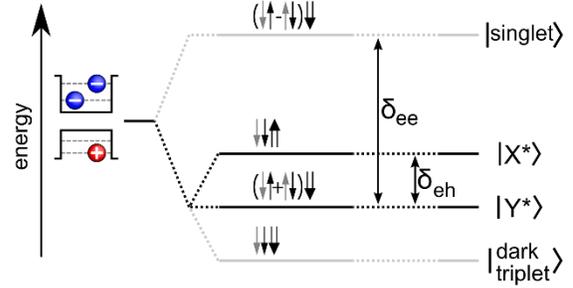

FIG.S 6: Possible spin configurations of the $p$-shell trions. Black (gray) ↓/↑ represent the spins of the $p$-shell ($s$-shell) electron, while ⇓/⇑ represent the heavy-hole (HH) spins, respectively. Each level has a two-fold Kramers degeneracy; the spin-inverted states are not shown explicitly. Effective electron-electron and electron hole exchange couplings are represented by $\delta_{ee}$ and $\delta_{eh}$, respectively.

# Derivation of Eq. (1)

Our procedure to calculate the differential transmission signal of the QD (see Eq. (1)) follows the treatment of the Maxwell-Liouville equations in Ref. [9]. Accordingly, we separate the calculation into two steps: First, the dynamics of the electronic system is calculated via a Liouville equation, including the impact of the in-coming light field as an external potential. In a second step, the dynamics of the electronic state causes a polarization field, which is used as a source for the macroscopic Maxwell equations to calculate the transmitted electric field and thereby the measured signal.

## Optically induced dynamics of the electronic system

In the first step, the dynamics of the electronic system with respect to the in-coming light field and relaxation processes is calculated. Therefore, we use the quantum mechanical Liouville equation with Lindblad dissipator

$$\dot{\rho}_{m,n}(t) = -\frac{i}{\hbar}\left(\left[\hat{H}^0 + \hat{H}^\gamma(t), \hat{\rho}(t)\right]\right)_{m,n} + D_{m,n}(\hat{\rho}). \quad (S\ 3)$$

$\rho_{m,n}$ describes the density matrix, $\hat{H}^0$ the stationary energies of the levels, $\hat{H}^\gamma$ the influence of the in-coming light field and $D_{m,n}(\hat{\rho})$ the Lindblad dissipator. The equation is evaluated in the basis built of the states depicted in FIG. 1(b) of the main paper.

In the following, we describe some approximations which lead to an analytical solution of Eq. (S 3). The validity of these approximations was checked by a numerical treatment of Eq. (S 3), which revealed no noticeable difference in any analyzed parameter set. For the analytical solution, we separate the dynamics into the following temporal steps: 1. Starting point is a complete occupation of |QDGS⟩, 2. Dynamics during the pump pulse, 3. Relaxation of the hole, 4. Dynamics between the incidence of the light pulses, 5. Dynamics during the probe pulse, 6. Dynamics after the probe pulse. This temporal separation is based on the different time scales of the processes involved and allows for a separate treatment of relaxation processes and light-matter interaction:

## Light-matter interaction - dynamics during the laser pulses

The light-matter interaction is described in the typical dipole and rotating wave approximation $\hat{H}^\gamma(t) = -\vec{E}_{\text{in}}^{(+)}(t) \cdot \hat{\vec{d}}^+ - \vec{E}_{\text{in}}^{(-)}(t) \cdot \hat{\vec{d}}$. The dipole operator is separated into the operators $\hat{\vec{d}}$ and $\hat{\vec{d}}^+$ describing relaxations from higher to lower and excitations from lower to higher energetic states, respectively. The in-coming electric field is separated into positive (+) and negative (-) frequency components with $\vec{E}_{\text{in}}^{(\pm)} = \vec{E}_{\text{pump}}^{(\pm)} + \vec{E}_{\text{probe}}^{(\pm)}$ and

$$\vec{E}_{\text{pulse}}^{(+)}(t) = E_{\text{pulse}}^0(t)e^{-i\varphi_{\text{pulse}}}e^{-i\omega_{\text{pulse}}t}\left(e_{\text{pulse}}^x \vec{e}_x + e_{\text{pulse}}^y \vec{e}_y\right). \quad (S\ 4)$$

Here, "pulse" refers to either "pump" or "probe", $E_{\text{pulse}}^0(t)$ defines the envelope, $\varphi_{\text{pulse}}$ the phase and $e_{\text{pulse}}^x \vec{e}_x + e_{\text{pulse}}^y \vec{e}_y$ denotes the polarization of the respective pulse. The frequencies $\omega_{\text{pump}}$ ($\omega_{\text{probe}}$) are tuned to an energy at the average between the |QDGS⟩ → |Y*⟩ and |QDGS⟩ → |X*⟩ (|X⟩ → |CBGS⟩ and |Y⟩ → |CBGS⟩ ) transitions. They are energetically sufficiently broad to cover both transitions but also narrow enough to provide a full separation between pump and probe transitions.

The calculation of the dynamics during a laser pulse is based on the idea, that relaxation and dephasing can be neglected during the pulses. Within the interaction picture using $\rho_{m,n}(t) = e^{-i\omega_{m,n}t}\tilde{\rho}_{m,n}(t)$ with $\omega_{m,n} = \frac{1}{\hbar}(H_{m,m}^0 - H_{n,n}^0)$, the remaining equation of motion reads

$$\dot{\tilde{\rho}}_{m,n}(t) = -\frac{i}{\hbar}\sum_j \left(\tilde{H}_{m,j}^{\gamma,\text{pulse}}(t)\tilde{\rho}_{j,n}(t) - \tilde{\rho}_{m,j}(t)\tilde{H}_{j,n}^{\gamma,\text{pulse}}(t)\right) \quad (S\ 5)$$

with

$$\widetilde{H}_{j,n}^{\gamma,\text{pulse}}(t) = H_{j,n}^{\gamma,\text{pulse}}(t)e^{i\omega_{j,n}t} = -E_{\text{pulse}}^0(t)e^{-i\varphi_{\text{pulse}}}e^{-i(\omega_{\text{pulse}}-\omega_{j,n})t}\left(e_{\text{pulse}}^x\vec{e}_x + e_{\text{pulse}}^y\vec{e}_y\right)\cdot\vec{d}_{j,n}^{\,+} + h.c. \quad \text{(S 6)}$$

and $h.c.$ denoting the hermitian adjoint term. Following the treatment of short laser pulses in Ref. [10], we can approximate the effect of the pulse by a jump condition of the density matrix and assign

$$\hat{\rho}^{\text{after pulse}} = e^{-\frac{i}{\hbar}\hat{\Lambda}^{\gamma,\text{pulse}}}\hat{\rho}^{\text{before pulse}}\left(e^{-\frac{i}{\hbar}\hat{\Lambda}^{\gamma,\text{pulse}}}\right)^{+} \quad \text{(S 7)}$$

with

$$\Lambda_{j,n}^{\gamma,\text{pulse}} = -A_{\text{pulse}}^0 e^{-i\varphi_{\text{pulse}}}\left(e_{\text{pulse}}^x\vec{e}_x + e_{\text{pulse}}^y\vec{e}_y\right)\cdot\vec{d}_{j,n}^{\,+}\tilde{\delta}_{\omega_{\text{pulse}},\omega_{j,n}} + h.c. \quad \text{(S 8)}$$

with $A_{\text{pulse}}^0 = \int_{-\infty}^{\infty}E_{\text{pulse}}^0(t)\,dt$ and $\tilde{\delta}_{\omega_{\text{pulse}},\omega_{j,n}}$ equals zero except for $\tilde{\delta}_{\omega_{\text{pump}},\omega_{X^*/Y^*,\text{QDGS}}} = 1$ and $\tilde{\delta}_{\omega_{\text{probe}},\omega_{\text{CBGS,X/Y}}} = 1$. Assuming perfectly linearly polarized transitions with $\langle\text{QDGS}|\hat{\vec{d}}|X^*/Y^*\rangle = d_0\vec{e}_{x/y}$ and $\langle X/Y|\hat{\vec{d}}|\text{CBGS}\rangle = d_0\vec{e}_{x/y}$ containing a real constant $d_0$, an explicit form of the non-vanishing matrix elements of $\hat{\Lambda}^{\gamma,\text{pulse}}$ is given by

$$\hat{\Lambda}^{\gamma,\text{pulse}} = -A_{\text{pulse}}^0\begin{pmatrix} & |\text{QDGS}\rangle & |Y^*\rangle & |X^*\rangle \\ & 0 & e^{i\varphi_{\text{pulse}}}\mu_y^* & e^{i\varphi_{\text{pulse}}}\mu_x^* \\ & e^{-i\varphi_{\text{pulse}}}\mu_y & 0 & 0 \\ & e^{-i\varphi_{\text{pulse}}}\mu_x & 0 & 0 \end{pmatrix} \quad \text{(S 9)}$$

$$\hat{\Lambda}^{\gamma,\text{pulse}} = -A_{\text{pulse}}^0\begin{pmatrix} & |\text{TGS}\rangle & |Y\rangle & |X\rangle & |\text{CBGS}\rangle \\ & 0 & 0 & 0 & 0 \\ & 0 & 0 & 0 & e^{i\varphi_{\text{pulse}}}\nu_y^* \\ & 0 & 0 & 0 & e^{i\varphi_{\text{pulse}}}\nu_x^* \\ & 0 & e^{-i\varphi_{\text{pulse}}}\nu_y & e^{-i\varphi_{\text{pulse}}}\nu_x & 0 \end{pmatrix} \quad \text{(S 10)}$$

with $\mu_x = e_{\text{pump}}^x d_0$, $\mu_y = e_{\text{pump}}^y d_0$, $\nu_x = e_{\text{probe}}^x d_0$ and $\nu_y = e_{\text{probe}}^y d_0$.

### Lindblad dissipator - dynamics between and after the laser pulses

The Lindblad dissipator is based on the idea of an interaction between a system (here the electronic system) and a bath (here the phonons) described by the Hamiltonian $\widehat{H}_{\text{SB}} = \sum_j \hat{A}_j \otimes \hat{B}_j$, with $\hat{A}_j$ acting on the system and $\hat{B}_j$ on the bath. Thereby the Lindblad dissipator can be expressed via [11]

$$\widehat{D}(\hat{\rho}) = \sum_{j,k}\gamma_{j,k}\left(\hat{A}_k\hat{\rho}\hat{A}_j^+ - \frac{1}{2}\left(\hat{A}_j^+\hat{A}_k\hat{\rho} + \hat{\rho}\hat{A}_j^+\hat{A}_k\right)\right) \quad \text{(S 11)}$$

with $\gamma_{j,k} = \Gamma_{j,k} + \Gamma_{k,j}^*$ and $\Gamma_{j,k} = \frac{1}{\hbar^2}\text{Tr}_B\left(\hat{B}_j^\dagger \hat{B}_k \hat{\rho}_B\right)$. $\text{Tr}_B$ defines the trace over the bath. One can separate the terms into two types with $\hat{D}(\hat{\rho}) = \hat{D}^{\text{relax.}}(\hat{\rho}) + \hat{D}^{\text{deph.}}(\hat{\rho})$:

1. $\hat{D}^{\text{relax.}}(\hat{\rho})$ describes relaxation processes as well as the corresponding dephasing/coherence transfer. Each operator $\hat{A}_j$ in Eq. (S 11) describes a transition of the form $\hat{A}_j = |f_j\rangle\langle i_j|$ from an initial state $|i_j\rangle$ to a final state $|f_j\rangle$. Within the present experiment, two relaxation processes take place: The relaxation of the hole (within $\tau_H \approx 390$ fs) and the subsequent relaxation of the electron (within $\tau_E \approx 85$ ps). Since these processes occur on different time scales, we can consider them separately.

   The relaxation of the hole can be described by the two operators $\hat{A}_{X^*} = |X\rangle\langle X^*|$ and $\hat{A}_{Y^*} = |Y\rangle\langle Y^*|$. The occupation is transferred via $D_{X^*,X^*}^{\text{relax.}} = -D_{X,X}^{\text{relax.}} = -\gamma_{X^*,X^*}\rho_{X^*,X^*}$ and $D_{Y^*,Y^*}^{\text{relax.}} = -D_{Y,Y}^{\text{relax.}} = -\gamma_{Y^*,Y^*}\rho_{Y^*,Y^*}$. The phase between $|X^*\rangle$ and $|Y^*\rangle$ declines with $D_{X^*,Y^*}^{\text{relax.}} = -\frac{1}{2}(\gamma_{X^*,X^*} + \gamma_{Y^*,Y^*})\rho_{X^*,Y^*}$, while the phase between $|X\rangle$ and $|Y\rangle$ grows with $D_{X,Y}^{\text{relax.}} = \gamma_{Y^*,X^*}\rho_{X^*,Y^*}$. Considering the microscopic form of the electron-phonon interaction [12], the facts that $|X^*\rangle$ and $|Y^*\rangle$ ($|X\rangle$ and $|Y\rangle$) have the same orbital wave function but just differ in their spin configuration, that the spin configuration is not changed during the relaxation from $|X^*\rangle$ to $|X\rangle$ ($|Y^*\rangle$ to $|Y\rangle$) and that the two relaxations $|X^*\rangle \to |X\rangle$ and $|Y^*\rangle \to |Y\rangle$ have to overcome the same energy, lead to the inference that $\hat{B}_{X^*} = \hat{B}_{Y^*}$. It follows that $\gamma_{Y^*,X^*} = \gamma_{X^*,X^*} = \gamma_{Y^*,Y^*}$ and consequently $D_{X,Y}^{\text{relax.}} = -D_{X^*,Y^*}^{\text{relax.}}$. Thus, the phase between $|X^*\rangle$ and $|Y^*\rangle$ is transferred to the phase between $|X\rangle$ and $|Y\rangle$ without loss of coherence. We name $\gamma_{Y^*,X^*} = \gamma_{X^*,X^*} = \gamma_{Y^*,Y^*} =: \frac{1}{\tau_H}$.

   The relaxation of the electron can be described by the two operators $\hat{A}_X = |\text{TGS}\rangle\langle X|$ and $\hat{A}_Y = |\text{TGS}\rangle\langle Y|$. The occupation is transferred via $D_{X,X}^{\text{relax.}} = -\gamma_{X,X}\rho_{X,X} - \frac{1}{2}(\gamma_{X,Y}\rho_{X,Y} + \gamma_{Y,X}\rho_{Y,X})$, $D_{Y,Y}^{\text{relax.}} = -\gamma_{Y,Y}\rho_{Y,Y} - \frac{1}{2}(\gamma_{X,Y}\rho_{X,Y} + \gamma_{Y,X}\rho_{Y,X})$ and $D_{TGS,TGS}^{\text{relax.}} = \gamma_{X,X}\rho_{X,X} + \gamma_{Y,Y}\rho_{Y,Y} + \gamma_{X,Y}\rho_{X,Y} + \gamma_{Y,X}\rho_{Y,X}$. The phase between $|X\rangle$ and $|Y\rangle$ declines with $D_{X,Y}^{\text{relax.}} = -\frac{1}{2}(\gamma_{X,X} + \gamma_{Y,Y})\rho_{X,Y} - \frac{1}{2}\gamma_{X,Y}(\rho_{X,X} + \rho_{Y,Y})$. Because the two transitions $|X\rangle \to |\text{TGS}\rangle$ and $|Y\rangle \to |\text{TGS}\rangle$ have different energies, the relative phase $\rho_{X,Y}$ oscillates, here with a period of $\sim 7.5$ ps, which is much faster than the relaxation time of $\tau_E \approx 85$ ps. Consequently, the gray-marked contributions vanish in time-average and are neglected in the following. Considering the electron-phonon interaction, both states $|X\rangle$ and $|Y\rangle$ have the same orbital wave function. However, the spin contributions change during the transition into $|\text{TGS}\rangle$. In fact, just the singlet-like part of $|X\rangle$ and $|Y\rangle$ can relax. As discussed in Ref. [2] and the previous chapter of this supplementary, the singlet is coupled to the energetically lower bright triplet state via $\delta_{ee}$. Thus, without $\delta_{eh}$, just the lower bright triplet state has a singlet-like part and can relax. However, the experimental data show a high degree of linear polarization of $XX_X^-$ and $XX_Y^-$ and thus a nearly complete mixture of the two bright triplet states through a strong $\delta_{eh}$. Thus, both bright triplet states have similar singlet-like parts and thereby we get $\gamma_{X,X} \approx \gamma_{Y,Y}$. A similar relaxation rate of $|X\rangle$ and $|Y\rangle$ is also verified by the pump-probe measurements. Throughout the paper we use $\gamma_{X,X} = \gamma_{Y,Y} =: \frac{1}{\tau_E} = \frac{1}{\tau_{XY}}$.

2. $\hat{D}^{\text{deph.}}(\hat{\rho})$ describes pure dephasing, which is caused by operators $\hat{A}_n = |n\rangle\langle n|$ with $n \in \{\text{QDGS}, Y^*, X^*, \text{TGS}, Y, X, \text{CBGS}\}$. In this case, the dissipator can be written as $D_{j,k}^{\text{deph.}} = \left(\gamma_{k,j} - \frac{1}{2}(\gamma_{j,j} + \gamma_{k,k})\right)\rho_{j,k}$.

Considering the phase between $|X^*\rangle$ and $|Y^*\rangle$, the pure dephasing is given by $D_{X^*,Y^*}^{\text{deph.}} = \left(\gamma_{Y^*,X^*} - \frac{1}{2}(\gamma_{X^*,X^*} + \gamma_{Y^*,Y^*})\right)\rho_{X^*,Y^*}$. Considering the fact, that the electron-phonon interaction does not act on the spin configuration, we know that $\hat{B}_{X^*} = \hat{B}_{Y^*}$, thereby $\gamma_{Y^*,X^*} = \gamma_{X^*,X^*} = \gamma_{Y^*,Y^*}$ and thus $D_{X^*,Y^*}^{\text{deph.}} = 0$. This shows, that the phase between $|X^*\rangle$ and $|Y^*\rangle$ is protected from pure dephasing. Same arguments show that the phase between $|X\rangle$ and $|Y\rangle$ is protected from pure dephasing as well.

Due to the fast relaxation of the hole and the nature of the occupation relaxation mechanisms, the only relevant pure dephasing terms here occur for $D_{X,\text{CBGS}}^{\text{deph.}} = -\frac{1}{\tau_\delta}\rho_{X,\text{CBGS}}$ and $D_{Y,\text{CBGS}}^{\text{deph.}} = -\frac{1}{\tau_\delta}\rho_{Y,\text{CBGS}}$.

Between the light pulses $\hat{H}^\gamma(t)$ vanishes and thus, just the free evolution due to $\hat{H}^0$ and the Lindblad dissipator are relevant. We treat the first relaxation of the hole and the following relaxation and dephasing processes separately:

**Relaxation of the hole:**

Because $\tau_H$ is much smaller than the other relaxation and dephasing times, we treat the relaxation of the hole by a jump condition of the density matrix directly after the jump condition of the pump. Therefore, we transfer $\rho_{i,j} \to \rho_{k,l}$ with $i,j \in \{X^*, Y^*\}$ and $k,l \in \{X, Y\}$. This enables a separate treatment of the two sub-systems $\{\text{QDGS}, Y^*, X^*\}$ and $\{\text{TGS}, Y, X, \text{CBGS}\}$.

**Further relaxation processes and free dynamics between and after the laser pulses:**

For the free dynamics between and after the laser pulses, as well as for the remaining dephasing and relaxation processes, an analytical solution can be given in the block-diagonal form

$$\hat{\rho}(t) = \begin{pmatrix} 0 & 0 \\ 0 & \hat{\rho}_{\text{measure}}(t) \end{pmatrix} \quad \text{(S 12)}$$

with

$$\hat{\rho}_{\text{measure}}(t) =$$

$$\begin{pmatrix} 1 - \rho_{Y,Y}(t_0)e^{-\frac{t}{\tau_E}} - \rho_{X,X}(t_0)e^{-\frac{t}{\tau_E}} & 0 & 0 & 0 \\ 0 & \rho_{Y,Y}(t_0)e^{-\frac{t}{\tau_E}} & \rho_{Y,X}(t_0)e^{i\omega_{XY}t}e^{-\frac{t}{\tau_{XY}}} & \rho_{Y,\text{CBGS}}(t_0)e^{i\omega_Y t}e^{-\frac{t}{\tau_\delta}} \\ 0 & \rho_{X,Y}(t_0)e^{-i\omega_{XY}t}e^{-\frac{t}{\tau_{XY}}} & \rho_{X,X}(t_0)e^{-\frac{t}{\tau_E}} & \rho_{X,\text{CBGS}}(t_0)e^{i\omega_X t}e^{-\frac{t}{\tau_\delta}} \\ 0 & \rho_{\text{CBGS},Y}(t_0)e^{-i\omega_Y t}e^{-\frac{t}{\tau_\delta}} & \rho_{\text{CBGS},X}(t_0)e^{-i\omega_X t}e^{-\frac{t}{\tau_\delta}} & \rho_{\text{CBGS},\text{CBGS}}(t_0) \end{pmatrix} \quad \text{(S 13)}$$

with the column labels $|\text{TGS}\rangle$, $|Y\rangle$, $|X\rangle$, $|\text{CBGS}\rangle$, and the time of the respective previous laser pulse $t_0$.

## Optical response of the QD and differential transmission signal

With the knowledge of the time-dependent density matrix, the polarization can be calculated via $\vec{P}(t) = \text{Tr}(\hat{\vec{d}}\hat{\rho}(t))$ with Tr defining the trace. Using $\vec{P}(t)$ as a source for the macroscopic Maxwell equations, the total electric field $\vec{E}_{\text{out}}(t)$ after the QD can be calculated. Following Ref. [9], we assume a thin QD sample with $z \in [0, l]$ and a monochromatic light field with frequency $\omega_c$, which results in $\vec{E}_{\text{out}}(t) = \vec{E}_{\text{in}}(t) + i\frac{\omega_c \mu_0 cl}{2}\vec{P}(t)$ with the speed of light $c$ and the vacuum permeability $\mu_0$. In frequency space, we get $\vec{E}_{\text{out}}(\omega) = \vec{E}_{\text{in}}(\omega) + i\frac{\omega_c \mu_0 cl}{2}\vec{P}(\omega)$ and thereby an intensity of the field following the QD of

$$I(\omega) \sim \left|\vec{E}_{\text{in}}(\omega) + i\frac{\omega_c \mu_0 cl}{2}\vec{P}(\omega)\right|^2 = \left|\vec{E}_{\text{in}}(\omega)\right|^2 + 2\frac{\omega_c \mu_0 cl}{2}\text{Im}\left(\vec{E}_{\text{in}}(\omega)\vec{P}^*(\omega)\right) + \left|\frac{\omega_c \mu_0 cl}{2}\vec{P}(\omega)\right|^2 \quad \text{(S 14)}$$

with Im denoting the imaginary part. The term $\left|\vec{E}_{\text{in}}(\omega)\right|^2$ is removed by detecting differential transmission signals. The polarization-induced light field is much smaller than the probe field $\left|\frac{\omega_c \mu_0 cl}{2}\vec{P}(\omega)\right| \ll \left|\vec{E}_{\text{in}}(\omega)\right|$. Thus, we can describe the differential transmission signal to a good approximation as $\left(\frac{\Delta T}{T}\right)(\omega) \sim \text{Im}\left(\vec{E}_{\text{in}}(\omega)\vec{P}^*(\omega)\right)$. Since we are interested in signals, which are energetically close to the transitions $|Y\rangle \to |\text{CBGS}\rangle$ and $|X\rangle \to |\text{CBGS}\rangle$, we can restrict our consideration to $\left(\frac{\Delta T}{T}\right)_{X/Y}(\omega) \sim \text{Im}\left(\vec{E}_{\text{in}}(\omega)\vec{P}^*_{X/Y}(\omega)\right)$ with $\vec{P}_X(\omega) = \text{FT}(\rho_{\text{CBGS},X}(t)d_0\vec{e}_x)(\omega)$ (similar for $\vec{P}_Y$), where FT denotes the Fourier transformation. For a time delay $t_D$ between the laser pulses, the solution of this procedure is given by

$$\left(\frac{\Delta T}{T}\right)_Y(\omega) \sim \frac{\sin\left(\frac{A^0_{\text{pump}}}{2\hbar}\sqrt{|\mu_x|^2+|\mu_y|^2}\right)^2 \sin\left(\frac{A^0_{\text{probe}}}{2\hbar}\sqrt{|\nu_x|^2+|\nu_y|^2}\right)}{(|\mu_x|^2+|\mu_y|^2)(|\nu_x|^2+|\nu_y|^2)^{\frac{3}{2}}} \left\{\frac{-\frac{1}{\tau_\delta}}{\frac{1}{\tau_\delta^2}+(\omega_Y-\omega)^2}\left[|\nu_y|^2|\nu_x|^2\left(|\mu_y|^2-|\mu_x|^2\right)e^{-\frac{t_D}{\tau_E}} + \right.\right.$$

$$|\nu_y|^2 \cos\left(\frac{A^0_{\text{probe}}}{2\hbar}\sqrt{|\nu_x|^2+|\nu_y|^2}\right)\left(|\mu_y|^2|\nu_y|^2+|\mu_x|^2|\nu_x|^2\right)e^{-\frac{t_D}{\tau_E}} +$$

$$|\nu_y|^2 \cos\left(\frac{A^0_{\text{probe}}}{2\hbar}\sqrt{|\nu_x|^2+|\nu_y|^2}\right)\left(\mu_y^*\mu_x\nu_y^*\nu_x e^{i\omega_{XY}t_D}+\mu_y\mu_x^*\nu_y\nu_x^* e^{-i\omega_{XY}t_D}\right)e^{-\frac{t_D}{\tau_{XY}}} +$$

$$\frac{|\nu_x|^2-|\nu_y|^2}{2}\left(\mu_y^*\mu_x\nu_y^*\nu_x e^{i\omega_{XY}t_D}+\mu_y\mu_x^*\nu_y\nu_x^* e^{-i\omega_{XY}t_D}\right)e^{-\frac{t_D}{\tau_{XY}}}\right] +$$

$$\left.\frac{\omega_Y-\omega}{\frac{1}{\tau_\delta^2}+(\omega_Y-\omega)^2}\left[\frac{|\nu_x|^2+|\nu_y|^2}{2i}\left(\mu_y^*\mu_x\nu_y^*\nu_x e^{i\omega_{XY}t_D}-\mu_y\mu_x^*\nu_y\nu_x^* e^{-i\omega_{XY}t_D}\right)e^{-\frac{t_D}{\tau_{XY}}}\right]\right\} \quad \text{(S 15)}$$

and

$$\left(\frac{\Delta T}{T}\right)_X(\omega) \sim \frac{\sin\left(\frac{A^0_{\text{pump}}}{2\hbar}\sqrt{|\mu_x|^2+|\mu_y|^2}\right)^2 \sin\left(\frac{A^0_{\text{probe}}}{2\hbar}\sqrt{|\nu_x|^2+|\nu_y|^2}\right)}{(|\mu_x|^2+|\mu_y|^2)(|\nu_x|^2+|\nu_y|^2)^{\frac{3}{2}}}\left\{\frac{-\frac{1}{\tau_\delta}}{\frac{1}{\tau_\delta^2}+(\omega_X-\omega)^2}\left[|\nu_y|^2|\nu_x|^2\left(|\mu_x|^2-|\mu_y|^2\right)e^{-\frac{t_D}{\tau_E}}+\right.\right.$$

$$|\nu_x|^2 \cos\left(\frac{A^0_{\text{probe}}}{2\hbar}\sqrt{|\nu_x|^2+|\nu_y|^2}\right)\left(|\mu_y|^2|\nu_y|^2+|\mu_x|^2|\nu_x|^2\right)e^{-\frac{t_D}{\tau_E}}+$$

$$|\nu_x|^2 \cos\left(\frac{A^0_{\text{probe}}}{2\hbar}\sqrt{|\nu_x|^2+|\nu_y|^2}\right)\left(\mu_y^*\mu_x\nu_y^*\nu_x e^{i\omega_{XY}t_D}+\mu_y\mu_x^*\nu_y\nu_x^*e^{-i\omega_{XY}t_D}\right)e^{-\frac{t_D}{\tau_{XY}}}+$$

$$\frac{|\nu_y|^2-|\nu_x|^2}{2}\left(\mu_y^*\mu_x\nu_y^*\nu_x e^{i\omega_{XY}t_D}+\mu_y\mu_x^*\nu_y\nu_x^*e^{-i\omega_{XY}t_D}\right)e^{-\frac{t_D}{\tau_{XY}}}\Bigg] -$$

$$\frac{\omega_X-\omega}{\frac{1}{\tau_\delta^2}+(\omega_X-\omega)^2}\left[\frac{|\nu_x|^2+|\nu_y|^2}{2i}\left(\mu_y^*\mu_x\nu_y^*\nu_x e^{i\omega_{XY}t_D}-\mu_y\mu_x^*\nu_y\nu_x^*e^{-i\omega_{XY}t_D}\right)e^{-\frac{t_D}{\tau_{XY}}}\right]\Bigg\} \quad \text{(S 16)}$$

Here we marked the terms leading to the absorptive (dispersive) line shape in blue (red) and the terms describing the $t_D$-dependence in green. For the cases discussed in the main paper, this solution can be simplified via the assumptions $|\mu_x|=|\mu_y|=:|\mu|$, $|\nu_x|=|\nu_y|=:|\nu|$, $\mu_y^*\mu_x\nu_y^*\nu_x=|\mu|^2|\nu|^2 e^{i\vartheta}$ and $\frac{A^0_{\text{probe}}}{\sqrt{2}\hbar}|\nu| \ll 1$, leading directly to Eq. (1).

## Spectro-Temporal Line Shape

Based on the derivation presented above, we can understand the spectro-temporal line shape observed in FIG. 2. Therefore we consider the susceptibility $\chi(\omega)$ defined by $\vec{P}(\omega)=\chi(\omega)\vec{E}_{\text{probe}}(\omega)$ with the polarization $\vec{P}(\omega)$ and the probe field $\vec{E}_{\text{probe}}(\omega)$. As usual, the imaginary part of the susceptibility describes absorption/emission and is visible in the differential transmission via $\left(\frac{\Delta T}{T}\right)(\omega) \sim \text{Im}\left(\vec{E}_{\text{probe}}(\omega)\vec{P}^*(\omega)\right) = -\left|\vec{E}_{\text{probe}}(\omega)\right|^2 \text{Im}(\chi(\omega))$. The real part of $\chi(\omega)$ defines a phase shift of the field and causes dispersion but is not visible in standard intensity measurements.

In the linear response regime (for example without pump pulse) the susceptibility has the well-known form $\chi_0(\omega)=\frac{(\omega_0-\omega)+i\gamma}{\gamma^2+(\omega_0-\omega)^2}$ with dephasing rate $\gamma$ and center frequency $\omega_0$, giving rise to a Lorentzian-shaped imaginary part describing the absorption and a "dispersion-shaped" real part.

In our case, the probe excites |CBGS⟩ from |X⟩ and |Y⟩, imprinting the phase of both |X⟩ and |Y⟩ on |CBGS⟩. Considering for example the polarization built by the relative phase between |X⟩ and |CBGS⟩ (XX$_X^-$), we get two terms: One term between |X⟩ and the part of |CBGS⟩ with a phase of |X⟩ imprinted, where the phases of |X⟩ cancel. This term describes the typical delay-independent Lorentzian shape of the intensity, plotted in FIG. 2(a). The second term arises between |X⟩ and the part of |CBGS⟩ with a phase of |Y⟩ imprinted, thus containing the relative phase between |X⟩ and |Y⟩ at $t_D$ via a phase factor $e^{-i\omega_{XY}t_D}$. When calculating $\text{Im}\left(e^{-i\omega_{XY}t_D}\chi_0(\omega)\right)$, depending on the delay time $t_D$, both Lorentzian-shaped and dispersion-shaped contributions show up, as is seen in FIG. 2(b) and (c). Thus, this second way to build up a polarization causes the quantum beats with the observed non-Lorentzian line shape.

# References:

[1] S. Mahapatra, K. Brunner, and C. Bougerol, Appl. Phys. Lett. **91**, 153110 (2007).

[2] C. Hinz, P. Gumbsheimer, C. Traum, M. Holtkemper, B. Bauer, J. Haase, S. Mahapatra, A. Frey, K. Brunner, D. E. Reiter, T. Kuhn, D. V. Seletskiy, and A. Leitenstorfer, Phys. Rev. B **97**, 045302 (2018).

[3] C. Genet and T. W. Ebbesen, Nature **445**, 39 (2007).

[4] P. W. Milonni and J. H. Eberly, *Laser Physics* (John Wiley & Sons, Inc., Hoboken, NJ, USA, 2010).

[5] I. A. Akimov, T. Flissikowski, A. Hundt, and F. Henneberger, Phys. Status Solidi **201**, 412 (2004).

[6] J. Huneke, I. D'Amico, P. Machnikowski, T. Thomay, R. Bratschitsch, A. Leitenstorfer, and T. Kuhn, Phys. Rev. B **84**, 115320 (2011).

[7] M. Holtkemper, D. E. Reiter, and T. Kuhn, Phys. Rev. B **97**, 075308 (2018).

[8] C. Traum, P. Henzler, S. Lohner, H. Becker, D. Nabben, P. Gumbsheimer, C. Hinz, J. F. Lippmann, S. Mahapatra, K. Brunner, D. V. Seletskiy, and A. Leitenstorfer, Rev. Sci. Instrum. **90**, 123003 (2019).

[9] S. Mukamel, *Principles of Nonlinear Optical Spectroscopy* (Oxford University Press, 1995).

[10] V. M. Axt, T. Kuhn, A. Vagov, and F. M. Peeters, Phys. Rev. B **72**, 125309 (2005).

[11] H.-P. Breuer and F. (Francesco) Petruccione, *The Theory of Open Quantum Systems* (Oxford University Press, 2002).

[12] G. D. Mahan, *Many-Particle Physics* (Plenum Press - New York and London, 1981).